\documentclass[12pt]{article}
\usepackage{epsfig}

\def\beq{\begin{equation}}
\def\eeq#1{\label{#1}\end{equation}}
\def\eeqn{\end{equation}}

\def\beqa{\begin{eqnarray}}
\def\eeqa#1{\label{#1}\end{eqnarray}}
\def\eeqan{\end{eqnarray}}

\let\bar=\overbar

\def\Dslash{\not{\hbox{\kern-4pt $D$}}}
\def\dslash{\not{\hbox{\kern-2pt $\del$}}}

\def\msb{{\bar{\ssstyle M \kern -1pt S}}}

\def\Title#1{\begin{center} {\Large {\bf #1} } \end{center}}

\begin{document}

\Title{Theory News on \boldmath$B_{s(d)} \to \mu^+\mu^-$\unboldmath\ Decays}

\bigskip\bigskip


\begin{raggedright}  

{\it Robert Fleischer\index{Fleischer, R.}\\
Nikhef, Science Park 105, NL-1098 XG Amsterdam,  NETHERLANDS,\\
Department of Physics and Astronomy, Vrije Universiteit Amsterdam\\
NL-1081 HV Amsterdam, NETHERLANDS}
\bigskip\bigskip
\end{raggedright}

\noindent
{\it {\bf Abstract:}  The rare decays $B_{s(d)} \to \mu^+\mu^-$ play a key role for the testing
of the Standard Model. An overview of the most recent theoretical
predictions of the corresponding branching ratios is given, emphasizing that the sizable
decay width difference $\Delta\Gamma_s$ of the $B_s$-meson system affects 
the $B_s\to \mu^+\mu^-$ channel. As a consequence, 
the calculated Standard Model branching ratio has to be upscaled by about $10\%$ to 
BR$(B_s\to \mu^+\mu^-)=(3.54\pm0.30)\times 10^{-9}$. This prediction is the reference 
value for the comparison with the time-integrated experimental branching ratio, where 
LHCb has recently reported $(3.2^{+1.5}_{-1.2})\times10^{-9}$ corresponding to the first 
evidence for $B_s\to \mu^+\mu^-$. The $\Delta\Gamma_s$ effects have also to be included 
in the constraints on the parameter space of New-Physics models following from the 
experimental data. Furthermore, $\Delta\Gamma_s$ makes a new observable through 
the effective $B_s\to \mu^+\mu^-$ lifetime accessible, which probes New Physics 
in a way complementary to the branching ratio and adds an exciting new topic to the
agenda for the high-luminosity upgrade of the LHC.

\vspace*{0.7truecm}

\noindent 
Proceedings of CKM 2012, the 7th International Workshop on the CKM Unitarity Triangle, University of Cincinnati, USA, 28 September -- 2 October 2012}

\section{Introduction}
The rare $B_{s(d)}\to\mu^+\mu^-$  decays receive only loop contributions from box 
and penguin topologies in the Standard Model (SM). Moreover, as only leptons are 
present in the final states, the hadronic sectors are very simple, involving the 
$B_{s(d)}$ decay constants $f_{B_{s(d)}}$. These transitions with their strongly suppressed 
rates offer hence powerful probes to search for footprints of ``New Physics" (NP) originating 
from beyond the SM (see \cite{BG} and references therein).

The most recent theoretical updates of the branching ratios read as follows \cite{BGGI}: 
\begin{eqnarray}
\mbox{BR}(B_s\to \mu^+\mu^-)_{\rm SM}&=&(3.23\pm0.27)\times 10^{-9}\label{BRs-SM}\\
\mbox{BR}(B_d\to \mu^+\mu^-)_{\rm SM}&=&(1.07\pm0.10)\times 10^{-10}\label{BRd-SM},
\end{eqnarray}
where the errors are dominated by those of lattice QCD determinations of the non-perturbative
$f_{B_{s(d)}}$ decay constants \cite{gamiz}. 

The limiting factor of the measurement of the 
$B_s\to \mu^+\mu^-$ branching ratio at hadron colliders is also due to a non-perturbative
quantity, the ratio $f_s/f_d$ of the fragmentation functions $f_q$ describing the probability 
that a $b$ quark fragments in a $\bar B^0_q$ meson. A new
method for determining $f_s/f_d$ using nonleptonic $\bar B^0_s\to D_s^+\pi^-$, 
$\bar B^0_d\to D^+K^-$, $\bar B^0_d\to D^+\pi^-$ decays \cite{FST,FST-fact} was recently 
implemented at LHCb \cite{LHCb-hadr}, resulting in good agreement with 
measurements using semileptonic decays \cite{LHCb-sl}. The $SU(3)$-breaking form-factor 
ratio entering the non-leptonic method has recently been calculated with lattice
QCD \cite{gamiz,FF-lat}.

In November 2012, LHCb has reported the first  evidence for $B^0_s\to \mu^+\mu^-$ 
at the $3.5\sigma$ level, with the following branching ratio \cite{LHCb-Bsmumu}:
\begin{equation}\label{LHCb-BR}
\mbox{BR}(B_s\to \mu^+\mu^-)=(3.2^{+1.5}_{-1.2})\times10^{-9},
\end{equation}
and the upper limit of $\mbox{BR}(B_d\to \mu^+\mu^-)<9.4 \times10^{-10}$ (95\% C.L.). These
results complement previous constraints from the CDF, D0, ATLAS and CMS collaborations 
discussed at this workshop (for a recent experimental review, see \cite{albrecht}). 

In spring 2012, LHCb announced another interesting result on a -- seemingly -- unrelated topic, 
a sizable width difference $\Delta\Gamma_s$ of the $B_s$-meson system \cite{LHCb-DGs}:
\begin{equation}\label{ys}
	y_s \equiv\frac{\Delta\Gamma_s}{2\,\Gamma_s}\equiv
	\frac{\Gamma_{\rm L}^{(s)} - \Gamma_{\rm H}^{(s)}}{2\,\Gamma_s}= 0.088 \pm 0.014,
\end{equation}
where $\Gamma_s$ is the inverse of the average $B_s$ lifetime $\tau_{B_s}$. 
A sizable value of $\Delta\Gamma_s$ was theoretically expected since decades (for a review,
see \cite{lenz}). 

The significant decay width difference $\Delta\Gamma_s$ leads to subtleties in the interpretation
of experimental data on $B_s$ decays in terms of branching ratios calculated from theory 
\cite{BR-paper} (see also \cite{DFN,DGMV}). This concerns, 
in particular, the $B_s\to\mu^+\mu^-$ decay
\cite{Bsmumu-paper}, where the impact of $\Delta\Gamma_s$ had so far not been taken 
into account in the relevant analyses. Moreover, this quantity offers another theoretically 
clean observable, the effective lifetime of the $B_s\to \mu^+\mu^-$ channel, 
which is complementary to the branching ratio.

In the case of the $B_d\to\mu^+\mu^-$ decay, this phenomenon is 
not of practical relevance as the decay width difference $\Delta\Gamma_d$ of the 
$B_d$-meson system has a tiny value of $\Delta\Gamma_d/\Gamma_d\sim10^{-3}$ 
in the SM \cite{lenz}.

\section{Observables}
Using the same notation as in  \cite{APS}, the low-energy effective Hamiltonian describing
the $\bar B^0_s\to\mu^+\mu^-$ decay can be written as
\begin{equation}\label{Heff}
{\cal H}_{\rm eff}=-\frac{G_{\rm F}}{\sqrt{2}\pi} \alpha V_{ts}^\ast V_{tb} 
\bigl[C_{10} O_{10} + C_{S} O_S + C_P O_P
 + C_{10}' O_{10}' + C_{S}' O_S' + C_P' O_P' \bigr].
\end{equation}
Here the short-distance physics is encoded in the Wilson coefficients $C_i$,  $C_i'$ of the
four-fermion operators
\begin{equation}
O_{10}=(\bar s \gamma_\mu P_L b) (\bar\ell\gamma^\mu \gamma_5\ell), \quad
O_S=m_b (\bar s P_R b)(\bar \ell \ell), \quad
O_P=m_b (\bar s P_R b)(\bar \ell \gamma_5 \ell),
\end{equation}
where $P_{L,R}\equiv(1\mp\gamma_5)/2$, $m_b$ is the $b$-quark mass, and the $O'_i$  
are obtained from the $O_i$ through the replacements $P_L \leftrightarrow P_R$.
The matrix elements can be expressed in terms of the $B_s$-meson decay constant 
$f_{B_s}$. Making the replacements $s\to d$ yields the Hamiltonian describing the 
$\bar B^0_d\to\mu^+\mu^-$ decay. 

Only the $O_{10}$ operator is present in the SM, with a real Wilson 
coefficient $C_{10}^{\rm SM}$ which governs the predictions in (\ref{BRs-SM}) and 
(\ref{BRd-SM}). An outstanding feature of $\bar B^0_s\to \mu^+\mu^-$ with respect to 
probing NP is the sensitivity to the (pseudo-)scalar lepton densities entering the 
$O_{(P)S}$ and $O_{(P)S}'$ operators, which have Wilson coefficients that are 
still largely unconstrained by the current data (see, for instance, \cite{APS}). 

The calculation of the $B_s\to\mu^+\mu^-$ observables is discussed in detail in
\cite{Bsmumu-paper}. Let us here first have a brief  
look at the following time-dependent rate asymmetry, which requires 
tagging information and knowledge of the muon helicity $\lambda$:
\begin{equation}\label{CP-hel}
\frac{\Gamma(B^0_s(t)\to \mu_\lambda^+\mu^-_\lambda)-
\Gamma(\bar B^0_s(t)\to \mu_\lambda^+
\mu^-_\lambda)}{\Gamma(B^0_s(t)\to \mu_\lambda^+\mu^-_\lambda)+
\Gamma(\bar B^0_s(t)\to \mu_\lambda^+\mu^-_\lambda)}
=\frac{C_\lambda\cos(\Delta M_st)+S_\lambda\sin(\Delta M_st)}{\cosh(y_st/\tau_{B_s}) + 
{\cal A}_{\Delta\Gamma}^\lambda \sinh(y_st/\tau_{B_s})}.
\end{equation}
The $y_s$ entering this expression was introduced in (\ref{ys}), $\Delta M_s$ is the 
mass difference between the heavy and light $B_s$ mass eigenstates, and 
\begin{equation}\label{C-lam}
\hspace*{-0.2truecm}C_\lambda\equiv\frac{1-|\xi_\lambda|^2}{1+|\xi_\lambda|^2}
=-\eta_\lambda\left[\frac{2|PS|\cos(\varphi_P-\varphi_S)}{|P|^2+|S|^2}  \right]
\end{equation}
\begin{equation}\label{S-lam}
\hspace*{-0.2truecm}S_\lambda\equiv \frac{2\,\mbox{Im}\,\xi_\lambda}{1+|\xi_\lambda|^2}
=\frac{|P|^2\sin2\varphi_P-|S|^2\sin 2\varphi_S}{|P|^2+|S|^2}
\end{equation}
\begin{equation}\label{ADG-lam}
\hspace*{-0.2truecm}
{\cal A}_{\Delta\Gamma}^\lambda\equiv \frac{2\,\mbox{Re}\,\xi_\lambda}{1+|\xi_\lambda|^2}
=\frac{|P|^2\cos 2\varphi_P-|S|^2\cos 2\varphi_S}{|P|^2+|S|^2}
\end{equation}
with 
\begin{equation}\label{P-expr}
P\equiv |P|e^{i\varphi_P}\equiv \frac{C_{10}-C_{10}'}{C_{10}^{\rm SM}}+\frac{M_{B_s} ^2}{2 m_\mu}
\left(\frac{m_b}{m_b+m_s}\right)\left(\frac{C_P-C_P'}{C_{10}^{\rm SM}}\right)
\end{equation}
\begin{equation}\label{S-expr}
S\equiv  |S|e^{i\varphi_S} \equiv \sqrt{1-4\frac{m_\mu^2}{M_{B_s}^2}}
\frac{M_{B_s} ^2}{2 m_\mu}\left(\frac{m_b}{m_b+m_s}\right)
\left(\frac{C_S-C_S'}{C_{10}^{\rm SM}}\right).
\end{equation}
The $P$ and $S$ with their CP-violating phases $\varphi_{P,S}$  
have been introduced such  that $P=1$ and $S=0$ in the SM. 
It should be noted that ${\cal S}_{\rm CP} \equiv S_\lambda$ and 
${\cal A}_{\Delta\Gamma}\equiv {\cal A}_{\Delta\Gamma}^\lambda$
do not depend on the muon helicity $\lambda$. 
In (\ref{S-lam}) and (\ref{ADG-lam}), the NP contribution to the $B^0_s$--$\bar B^0_s$ mixing 
phase $\phi_s=\phi_s^{\rm SM}+\phi_s^{\rm NP}$ was neglected, where the SM piece is given 
by  $\phi_s^{\rm SM}\equiv2\mbox{arg}(V_{ts}^\ast V_{tb})\approx -2^\circ$. 
This effect can straightforwardly 
be included through $2\varphi_{P,S}\to 2\varphi_{P,S}-\phi_s^{\rm NP}$.
The LHCb data for CP violation in $B_s\to J/\psi \phi, J/\psi f_0(980)$ already constrain 
$\phi_s^{\rm NP}$ to the few-degree level \cite{LHCb-DGs,LHCb-phis}, 
in contrast to the $\varphi_{P,S}$.  
Neglecting the impact of $\Delta\Gamma_s$, the CP asymmetries (\ref{CP-hel}) 
were considered for $B_{s,d}\to\ell^+\ell^-$ decays within various NP scenarios in the 
previous literature \cite{HL}--\cite{CKWW}.

As it is experimentally challenging to measure the muon helicity, we introduce  
\begin{equation}\label{rate-no-lam}
\Gamma({B}_s^0(t)\to \mu^+\mu^-)\equiv \sum_{\lambda={\rm L,R}}
\Gamma({B}_s^0(t)\to \mu^+_\lambda \mu^-_\lambda),
\end{equation}
and 
\begin{equation}\label{CP-no-lam}
\frac{\Gamma(B^0_s(t)\to \mu^+\mu^-)-
\Gamma(\bar B^0_s(t)\to \mu^+\mu^-)}{\Gamma(B^0_s(t)\to \mu^+\mu^-)+
\Gamma(\bar B^0_s(t)\to \mu^+\mu^-)}
=\frac{{\cal S}_{\rm CP}\sin(\Delta M_st)}{\cosh(y_st/ \tau_{B_s}) + 
{\cal A}_{\Delta\Gamma} \sinh(y_st/ \tau_{B_s})},
\end{equation}
where $\Gamma(\bar B^0_s(t)\to \mu^+\mu^-)$ is defined in analogy to (\ref{rate-no-lam}). 

A measurement of the CP asymmetry in (\ref{CP-no-lam}) would be most interesting 
as a non-vanishing value would immediately indicate new CP-violating phases, 
as illustrated recently in \cite{BDeFG}--\cite{BG-2}. Unfortunately, since tagging and time 
information are required, this is still an experimental challenge. An expression  analogous 
to (\ref{CP-no-lam}) holds also for $B_d\to\mu^+\mu^-$, where $y_d$ is negligibly small.

\section{Branching Ratio}
The first step for the experimental exploration of the $B_s\to\mu^+\mu^-$ decay 
is the extraction of a branching ratio from the untagged rate
\begin{equation}
	\langle \Gamma(B_s(t)\to \mu^+\mu^-)\rangle
	\equiv\ \Gamma(B^0_s(t)\to \mu^+\mu^-)+ \Gamma(\bar B^0_s(t)\to \mu^+\mu^-) 
\end{equation}
by ignoring the decay-time information \cite{BR-paper,DFN}:
\begin{equation}\label{BR-exp}
      {\rm BR}\left(B_s \to \mu^+\mu^-\right)_{\rm exp}
       \equiv \frac{1}{2}\int_0^\infty \langle \Gamma(B_s(t)\to \mu^+\mu^-)\rangle\, dt.
\end{equation}
The experimental branching ratio in (\ref{LHCb-BR}) refers to this branching ratio concept. 

On the other hand, theorists usually consider and calculate the following CP-averaged 
branching ratio:
\begin{equation}\label{BR-theo}
{\rm BR}\left(B_s \to \mu^+\mu^-\right)_{\rm theo}\equiv 
	\frac{\tau_{B_s}}{2}\langle \Gamma(B_s(t)\to \mu^+\mu^-)\rangle\Big|_{t=0},
\end{equation}
where the $B^0_s$--$\bar B^0_s$ oscillations are ``switched off" by choosing $t=0$. The
SM prediction in (\ref{BRs-SM}) actually refers to this branching ratio definition and satisfies
\begin{equation}
\frac{\mbox{BR}(B_s\to\mu^+\mu^-)_{\rm theo}}{\mbox{BR}(B_s\to\mu^+\mu^-)_{\rm SM}}=
|P|^2+|S|^2.
\end{equation}

As was pointed out in \cite{Bsmumu-paper}, 
the experimental branching ratio (\ref{BR-exp}) can be converted into the theoretical branching 
ratio (\ref{BR-theo}) with the help of the following expression:
\begin{equation}\label{BRratio}
        {\rm BR}(B_s \to \mu^+\mu^-)_{\rm theo}
        =   \left[\frac{1-y_s^2}{1 + {\cal A}_{\Delta\Gamma}\, y_s}\right] 
        {\rm BR}(B_s \to \mu^+\mu^-)_{\rm exp},
\end{equation}
where it is essential that ${\cal A}_{\Delta\Gamma}\equiv {\cal A}_{\Delta\Gamma}^\lambda$
does actually not depend on the muon helicity.

Looking at (\ref{ADG-lam})--(\ref{S-expr}), we observe that NP may affect 
${\cal A}_{\Delta\Gamma}$ through the  Wilson coefficients so that 
this observable is currently unknown. However, within the SM, we have the 
theoretically clean prediction of ${\cal A}_{\Delta\Gamma}^{\rm SM}=+1$. Using (\ref{BRratio}),  
we hence rescale the theoretical SM branching ratio in (\ref{BRs-SM}) by a factor of 
$1/(1-y_s)$, yielding 
\begin{equation}
\mbox{BR}(B_s\to \mu^+\mu^-)_{\rm SM}|_{y_s}=(3.54\pm0.30)\times 10^{-9}
\end{equation}
for the value of $y_s$ in (\ref{ys}). This is the SM reference value for the 
comparison with the experimental branching ratio (\ref{LHCb-BR}).

\section{Effective Lifetime}
In the future, once the $B_s\to\mu^+\mu^-$ signal has been well established and 
more data become available, also the decay-time information can be taken into account
in the experimental analysis. This will allow a measurement of the effective lifetime
\begin{equation}
\tau_{\mu^+\mu^-}\equiv \frac{\int_0^\infty t\,\langle \Gamma(B_s(t)\to \mu^+\mu^-)\rangle\, dt}
        {\int_0^\infty \langle \Gamma(B_s(t)\to \mu^+\mu^-)\rangle\, dt}
        = \frac{\tau_{B_s}}{1-y_s^2}\left[\frac{1+2\,{\cal A}_{\Delta\Gamma}y_s + y_s^2}
        {1 + {\cal A}_{\Delta\Gamma} y_s}\right],
\end{equation}
so that also the observable
\begin{equation}
 {\cal A}_{\Delta\Gamma}  = \frac{1}{y_s}\left[\frac{(1-y_s^2)\tau_{\mu^+\mu^-}-(1+
 y_s^2)\tau_{B_s}}{2\tau_{B_s}-(1-y_s^2)\tau_{\mu^+\mu^-}}\right]
\end{equation} 
can be extracted from the data \cite{BR-paper,Bsmumu-paper}. The effective lifetime allows 
the conversion of the experimental $B_s\to\mu^+\mu^-$ branching ratio into its theoretical 
counterpart through
\begin{equation}\label{BRmumu-correct}
{\rm BR}\left(B_s \to \mu^+\mu^-\right)_{\rm theo}
=\left[2 - \left(1-y_s^2\right)\frac{\tau_{\mu^+\mu^-}}{\tau_{B_s}}\right] 
{\rm BR}\left(B_s \to \mu^+\mu^-\right)_{\rm exp}.
\end{equation}
This relation holds no matter whether NP contributions are 
present in $B_s\to\mu^+\mu^-$ or whether this decay is fully governed by the SM.

The effective $B_s\to\mu^+\mu^-$ lifetime and the extraction of ${\cal A}_{\Delta\Gamma}$
from untagged data samples are exciting new aspects for the exploration of the
$B_s\to\mu^+\mu^-$ decay at the high-luminosity upgrade of the LHC. An extrapolation 
from current measurements of the effective $B_s\to J/\psi\,f_0(980)$ and 
$B_s\to K^+K^-$ lifetimes by the CDF and LHCb collaborations to 
$\tau_{\mu^+\mu^-}$ indicates that a precision of $5\%$ or better may be feasible 
\cite{Bsmumu-paper}. Detailed experimental studies are strongly encouraged.

\section{Constraints on New Physics}
The $\Delta\Gamma_s$ affects also the NP constraints that can be obtained
by comparing the experimental $B_s\to\mu^+\mu^-$ branching ratio information 
with the SM picture \cite{Bsmumu-paper}:
\begin{equation}\label{R-def}
R\equiv 
\frac{\mbox{BR}(B_s \to \mu^+\mu^-)_{\rm exp}}{\mbox{BR}(B_s \to \mu^+\mu^-)_{\rm SM}}=
\left[\frac{1+y_s\cos2\varphi_P}{1-y_s^2}  \right] |P|^2+
\left[\frac{1-y_s\cos2\varphi_S}{1-y_s^2}  \right] |S|^2.
\end{equation}
Using (\ref{BRs-SM}) and (\ref{LHCb-BR}) yields $R=1.0^{+0.5}_{-0.4}$, where the errors have
been added in quadrature.

The $R$ ratio can be converted into ellipses in the $|P|$--$|S|$ plane which depend on the
CP-violating phases $\varphi_{P,S}$. Since the latter quantities are unknown, $R$ fixes 
actually a circular band with the upper bounds $|P|, |S|\leq\sqrt{(1+y_s)R}$. As the
experimental information on $R$ does not allow us to separate the $S$ and $P$ contributions,
still significant NP contributions may be hiding in the $B_s\to\mu^+\mu^-$ channel.

This situation can be resolved by measuring the effective lifetime $\tau_{\mu^+\mu^-}$ and the associated ${\cal A}_{\Delta\Gamma}$ observable, since
\begin{equation}\label{S-ADG}
|S|=|P|\sqrt{\frac{\cos2\varphi_P-{\cal A}_{\Delta\Gamma}}{\cos2\varphi_S
+{\cal A}_{\Delta\Gamma}}}
\end{equation}
fixes a straight line through the origin in the $|P|$--$|S|$ plane. For illustrations, the
reader is referred to the figures shown in \cite{Bsmumu-paper}.

In the most recent analyses of the constraints on NP parameter space that are implied 
by the experimental upper bound on the $B_s\to\mu^+\mu^-$ branching ratio for various 
extensions of the SM, authors have now started to take the effect of $\Delta\Gamma_s$ 
into account (see, for instance, \cite{BDeFG}--\cite{BG-2}
and the papers in \cite{WCSY}--\cite{Nan}).

\section{Conclusions}\label{sec:concl}
We live in exciting times for analyses of rare leptonic $B$ decays. While the experimental
upper bound for the $B_d\to\mu^+\mu^-$ channel is still about an order of magnitude above
the SM expectation, LHCb has recently reported the first evidence for $B_s\to\mu^+\mu^-$
at the $3.5\sigma$ level, with a first experimental branching ratio of 
$(3.2^{+1.5}_{-1.2})\times10^{-9}$.

In view of the sizable decay width difference $\Delta\Gamma_s$ 
of the $B_s$-meson system, we have to deal with subtleties in the extraction of $B_s$ 
branching ratio information from the data, but we get also new observables, encoded in 
effective decay lifetimes. This is also the case for the $B_s\to\mu^+\mu^-$ decay, where
the theoretical SM branching ratio in (\ref{BRs-SM}) has to be rescaled by $1/(1-y_s)$ for 
the comparison with the experimental branching ratio, resulting in the SM reference value 
of $(3.54\pm0.30)\times 10^{-9}$. The sizable value of $\Delta\Gamma_s$ propagates also 
into the constraints on NP parameters that can be obtained from the $B_s\to\mu^+\mu^-$
branching ratio information, and can be included by means of the formulae presented
above. 

The future measurement of the effective $B_s\to\mu^+\mu^-$ lifetime  will allow 
the inclusion of the $\Delta\Gamma_s$ effect in the conversion of the experimental 
information into the theoretical branching ratio. Furthermore, the ${\cal A}_{\Delta\Gamma}$ 
encoded in $\tau_{\mu^+\mu^-}$ provides a new, theoretically clean NP probe that 
may still show large NP effects, in particular for scenarios with new contributions 
to the Wilson coefficients of the four-fermion operators with (pseudo-)scalar 
$\ell^+\ell^-$ densities. These features offer an exciting new topic for the next 
round of precision at the high-luminosity upgrade of the LHC. 

It will be most interesting to monitor the future experimental progress on the exploration of
$B_{s(d)}\to\mu^+\mu^-$ decays in this decade and how their increasingly more 
stringent constraints on the parameter space of specific NP models will complement 
other flavor probes and the direct searches for NP at the LHC.

\bigskip
{\it Acknowledgement:} \, I would like to thank my PhD students and colleagues 
for the enjoyable collaboration on topics discussed above.

\end{document}